\documentclass[onecolumn,showpacs]{revtex4}

\usepackage{graphicx}% Include figure files
\usepackage{dcolumn}% Align table columns on decimal point
\usepackage{bm}% bold math

\input epsf

\begin{document}

\title{\Large Generalized second law of thermodynamics in presence of interacting tachyonic field and scalar (phantom) field }

\author{\bf  Surajit
Chattopadhyay$^1$\footnote{surajit$_{_{-}}2008$@yahoo.co.in} and
Ujjal Debnath$^2$\footnote{ujjaldebnath@yahoo.com ,
ujjal@iucaa.ernet.in}}

\affiliation{$^1$Department of Computer Application, Pailan
College of Management and Technology, Bengal Pailan Park,
Kolkata-700 104, India.\\
$^2$Department of Mathematics, Bengal Engineering and Science
University, Shibpur, Howrah-711 103, India. }

\date{\today}

\begin{abstract}
In the present work, we have considered tachyonic field, phantom
field and scalar field in both interacting and non-interacting
situations and investigated the validity of the generalized second
law of thermodynamics in a flat FRW universe. We found that in all
cases, excepting the phantom field dominated universe, the
derivative of the entropy is remaining at negative level and is
increasing with the decrease in the redshift.
\end{abstract}


\maketitle

Since the discovery of the fact that presently there is an
accelerating expansion of the universe; different suggestions have
been made in order to explain the accelerated expansion. Among the
various suggestions, one is the hypothesis of dark energy (DE),
which is a fluid of negative pressure representing about $70\%$ of
the total energy of the universe [1, 2, 3, 4]. The distinguishing
feature of dark energy is that it violates the strong energy
condition $\rho+3p>0$ [5]. Strength of this acceleration depends
on the theoretical model employed while interpreting the data. A
wide range of scenarios have been proposed to explain this
acceleration but most of them cannot explain all the features of
the universe or they have so many parameters that they are
difficult to fit. The models which have been discussed widely in
literature are those which consider vacuum energy (cosmological
constant) as DE, introduce a fifth element and dub it quintessence
$(\omega>-1)$ or scenarios named phantom with $\omega < -1$, where
$\omega=p/\rho$ is a parameter of state [6].
\\\\
 One remarkable feature of the phantom
model is that the universe will end with a ``big rip" (future
singularity). That is, for phantom dominated universe, its total
lifetime is finite. Before the death of the universe, the phantom
dark energy will rip apart all bound structures like the Milky
Way, solar system, Earth and ultimately the molecules, atoms,
nuclei and nucleons of which we are composed. In the quintessence
model, the equation state $\omega=p/\rho$ is always in the range
$-1\leq \omega \leq 1$ for the potential $V(\phi)>0$. It has been
shown in the reference [7] that the decay of an unstable D-brane
produces pressure-less gas with finite energy density that
resembles classical dust. The cosmological effects of the tachyon
rolling down to its ground state have been discussed by the
reference [8]. There have been many works aimed at connecting the
string theory with inflation. While doing so, various ideas in
string theory based on the concept of branes have proved
themselves fruitful. Rolling tachyon matter associated with
unstable D-branes has an interesting equation of state which
smoothly interpolates between $-1$ and 0. As the Tachyon field
rolls down the hill, the universe experiences accelerated
expansion and at a particular epoch the scale factor passes
through the point of inflection marking the end of inflation [9].
\\\\
To obtain a suitable evolution of the Universe an interaction is
often assumed such that the decay rate should be proportional to
the present value of the Hubble parameter for good fit to the
expansion history of the Universe as determined by the Supernovae
and CMB data [10]. These kinds of models describe an energy flow
between the components so that no components are conserved
separately. There are several works on the interaction between
dark energy (tachyon or phantom) and dark matter [11, 12, 13, 14],
where phenomenologically introduced different forms of interaction
term.
\\\\
Izquierdo and Pavon [15] explored the thermodynamic consequences
of phantom-dominated universe and mentioned that one must take
into account that ever accelerating universe has a future event
horizon (or cosmological horizon). Since the horizon implies a
classically unsurmountable barrier to our ability to see what lies
beyond it, it is natural to attach an entropy to the horizon size
(i.e., to its area) for it is a measure of our ignorance about
what is going on in the other side [16]. Setare [16] investigated
the validity of the generalized second law of thermodynamics for
the quintom model of dark energy. Setare and Shafei [17] employed
a holographic model of dark energy to investigate the validity of
the first and second laws of thermodynamics in a non-flat (closed)
universe. Setare [18] considered the interacting holographic model
of dark energy to investigate the validity of the generalized
second law of thermodynamics in a non-flat (closed) universe
enclosed by the event horizon. The present work is different from
the works mentioned above. In the present work, we have
investigated the validity of the second law of thermodynamics in
the presence of interacting tachyonic field and scalar (phantom)
field in a spatially flat isotropic and homogeneous  FRW universe
whose metric is given by

\begin{equation}
ds^{2}=dt^{2}-a^{2}(t)\left[dr^{2}+r^{2}(d\theta^{2}+sin^{2}\theta
d\phi^{2})\right]
\end{equation}

where $a(t)$ is the scale factor. we consider a two fluid model
consisting of tachyonic field and scalar field (or phantom
field).
\\\\
Before going into the technical details we discuss the motivation
behind this study. In various previous works, the present
accelerated expansion of the universe has been studied by means of
single components of dark energy. Sami et al [19] examined the
possibility of rolling tachyon to play the dual role of inflaton
at early epochs and dark matter at late times. Chattopadhyay et al
[20] studied the acceleration of the universe in presence of
tachyonic field and showed that it interpolates between dust and
$\Lambda$CDM. Sing et al [21] studied the general features of the
dynamics of the phantom field in the cosmological context. In the
case of inverse coshyperbolic potential, they [21] demonstrated
that the phantom field can successfully drive the observed current
accelerated expansion of the universe with the equation of state
parameter $\omega<-1$. \\\\
 In the recent past, interacting dark
energy candidates has been studied by various authors. Cai and
Wang [12] studied a cosmological model in which phantom dark
energy is coupled to dark matter by phenomenologically introducing
a coupled term to the equations of motion of dark energy and dark
matter. Macorra [22] studied the cosmological evolution of two
coupled scalar fields with an arbitrary interaction term and
discussed the possibility of having one of the scalar fields as of
dark energy while the other could be a scalar field redshifting as
matter. An interaction between phantom field and tachyonic field
was considered by Chattopadhyay and Debnath [23]. Motivated by the
investigations on the validity of generalized second law of
thermodynamics in presence of phantom field [15], holographic dark
energy [17], quintom [16] and surveying the literatures on
two-component models [12, 18, 22, 23, 24, 25] we decided to
investigate the validity of the generalized second law of
thermodynamics in presence of interacting as well as
non-interacting tachyonic field and scalar (phantom) field.
\\\\
As we are considering a two-component model, the total energy
density and pressure are respectively given by
\begin{equation}
\rho_{tot}=\rho_{1}+\rho_{2}
\end {equation}
and
\begin{equation}
p_{tot}=p_{1}+p_{2}
\end {equation}

The energy density $\rho_{1}$ and pressure $p_{1}$ for tachyonic
field $\phi_{1}$ with potential $V_{1}(\phi_{1})$ are respectively
given by [23]

\begin{equation}
\rho_{1}=\frac{V_{1}(\phi_{1})}{\sqrt{1-{\dot{\phi}_{1}}^{2}}}
\end{equation}
and
\begin{equation}
p_{1}=-V_{1}(\phi_{1}) \sqrt{1-{\dot{\phi}_{1}}^{2}}
\end{equation}

The energy density $\rho_{2}$ and pressure $p_{2}$ for scalar
field (or phantom field) $\phi_{2}$ with potential
$V_{2}(\phi_{2})$ are respectively given by [23]

\begin{equation}
\rho_{2}=\frac{\epsilon}{2}~\dot{\phi}_{2}^{2}+V_{2}(\phi_{2})
\end{equation}
and
\begin{equation}
p_{2}=\frac{\epsilon}{2}~\dot{\phi}_{2}^{2}-V_{2}(\phi_{2})
\end{equation}

where, $\epsilon=1$ for scalar field and $\epsilon=-1$ for
phantom field.\\

Therefore, the conservation equation reduces to

\begin{equation}
\dot{\rho}_{1}+3H(\rho_{1}+p_{1})=-Q
\end{equation}
and
\begin{equation}
\dot{\rho}_{2}+3H(\rho_{2}+p_{2})=Q
\end{equation}

Here, $Q$ is the interaction term. If $Q$ is positive, then there
will be a transfer of energy from $\rho_{1}$ to $\rho_{2}$ [24].
For solving the above equations different forms for the
interaction term $Q$ are considered. For getting convenience while
integrating equation (8), we have chosen $Q=3\delta H\rho_{1}$
where $\delta$ is the interaction parameter and $H=\dot
a(t)/a(t)$. Similar choice of interaction term has been made in
[26]. A detailed discussion on $Q$ is available in [25].
\\\\
In an earlier work by the authors of the present work (Reference
[23]), the solutions for $\dot{\phi_{1}}$, $\dot{\phi_{2}}$,
$V_{1}$, and $V_{2}$ were obtained under interaction as follows:

\begin{equation}
\dot{\phi}_{1}^{2}=\left[-\delta+\left(\frac{c}{a^{3}}
\right)^{\frac{2(1+\delta)}{1+2m}}\right]\left[1+\left(\frac{c}{a^{3}}
\right)^{\frac{2(1+\delta)}{1+2m}}\right]^{-1}
\end{equation}

\begin{equation}
\dot{\phi}_{2}^{2}=-\frac{2\dot{H}}{\epsilon}+\frac{1}{\epsilon}\left[1+\left(\frac{c}{a^{3}}
\right)^{\frac{2(1+\delta)}{1+2m}}\right]^{m-\frac{1}{2}}
\left[\delta-\left(\frac{c}{a^{3}}
\right)^{\frac{2(1+\delta)}{1+2m}}\right](1+\delta)^{-m-\frac{1}{2}}
\end{equation}

\begin{equation}
V_{1}=\left[1+\left(\frac{c}{a^{3}}
\right)^{\frac{2(1+\delta)}{1+2m}} \right]^{m}(1+\delta)^{-m}
\end{equation}

\begin{equation}
V_{2}=\dot{H}+3H^{2}-\frac{1}{2}\left[1+\left(\frac{c}{a^{3}}
\right)^{\frac{2(1+\delta)}{1+2m}}\right]^{m-\frac{1}{2}}
\left[2+\delta+\left(\frac{c}{a^{3}}
\right)^{\frac{2(1+\delta)}{1+2m}}\right](1+\delta)^{-m-\frac{1}{2}}
\end{equation}

where, $c$ is an integration constant, $m>0$, $n>0$, and it was
assumed that $V_{1}=(1-\dot{\phi}_{1}^{2})^{-m}$, and
$V_{2}=n\dot{\phi}_{2}^{2}$.
\\\\
In the present work, we consider the universe as a flat FRW
universe and take into account that the accelerating universe has
a future event horizon $R_{h}$, which is also named as
cosmological horizon. The radius of observer's event horizon is
given by [16, 27]

\begin{equation}
R_{h}=a \int_{t}^{\infty}\frac{dt}{a}= a
\int_{a}^{\infty}\frac{da}{Ha^{2}}
\end{equation}

To study the generalized second law (GSL) through the universe
under the interaction between tachyonic field and scalar (phantom)
field we would examine the nature of the derivative of the normal
entropy $S$ in presence of interaction. It is a proven fact that
for phantom dominated universe $\dot{S}>0$ and for a quintessence
dominated universe $\dot{S}<0$ [16]. Our target is to answer the
question : \emph{Is $\dot{S}>0$ under the interaction between
tachyonic field and scalar (phantom) field}?.\\\\

We consider the FRW universe as a thermodynamical system with the
horizon surface as a boundary of the system. In general, the
radius of the event horizon $R_{h}$ is not constant but changes
with time. Let $dR_{h}$ be an infinitesimal change of the event
horizon radius during a time of interval $dt$. This small
displacement $dR_{h}$ will cause a small change $dV$ in the volume
$V$ of the event horizon. Each space–time describing a
thermodynamical system and satisfying Einstein's equations differs
infinitesimally in the extensive variables volume, energy and
entropy by $dV$ , $dE$ and $dS$, respectively, while having the
same values for the intensive variables temperature $T$ and
pressure $P$ [18]. Thus, for these two space–times describing two
thermodynamical states, there must exist some relation among these
thermodynamic quantities. It turns out that the differential form
of the Friedman equation can be rewritten as a universal form [16,
17]

\begin{equation}
TdS=dE+pdV=(p+\rho)dV+Vd\rho
\end{equation}

Also we know that

\begin{equation}
H^{2}=\frac{1}{3}\rho
\end{equation}
and
\begin{equation}
\dot{H}=-\frac{1}{2}(p+\rho)
\end{equation}

Using $V=\frac{4}{3}\pi R_{h}^{3}$ in equation (15) we get

\begin{equation}
TdS=-2\dot{H}dV+Vd\rho=-8\pi R_{h}^{2}\dot{H}dR_{h}+8\pi
R_{h}^{3}dH
\end{equation}

From equation (18), it can be obtained that

\begin{equation}
\dot{S}=\frac{8 \pi \dot{H}R_{h}^{2}}{T}
\end{equation}

If the horizon entropy is taken to be $S_{h}=\pi R_{h}^{2}$ [16],
we get

\begin{equation}
\dot{S}+\dot{S}_{h}=\frac{8 \pi \dot{H}R_{h}^{2}}{T}+2 \pi
R_{h}\dot{R}_{h}
\end{equation}

Taking the temperature $T=\frac{1}{2\pi R_{h}}$ and using
$\dot{H}R_{h}=HR_{h}-1$ [16] we can write

\begin{equation}
\dot{S}+\dot{S}_{h}=\dot{S}_{X}=16 \pi^{2}\dot{H}R_{h}^{3}+2 \pi
R_{h} (HR_{h}-1)
\end{equation}

At this stage it should be stated that the thermal equilibrium of
the different components with one another and/or with the horizon
cannot be assured. In this work it is acknowledged that the
thermal equilibrium is just an additional hypothesis.
\\

In the present problem, while considering tachyonic field, scalar
field and phantom field separately, the suffix $X$ of equation
(21) would be replaced by $tachyon$, $scalar$, and $phantom$
respectively. While considering the mixture of tachyon and scalar
(phantom) fields as a mixture without interaction (i.e.
$\delta=0$), $X$ would be replaced by $mixture$ and in the
interacting case (i.e. $\delta\neq 0$), $X$
would be replaced by $total$.\\

Equations (16) and (17) are now implemented with the pressure and
energy densities corresponding to the cases mentioned in the last
paragraph. Subsequently, $R_{h}$ could be obtained in all the
cases. Furthermore, $\dot{S}_{X}$ was obtained based on (16) and
(17). The $\dot{S}_{X}$ are plotted against the redshift
$z=1-a^{-1}$.
\\

Now we investigate the GSL of thermodynamics when the universe is
filled with (i) only tachyonic field, (ii) only phantom field,
(iii) only normal scalar field, (iv) mixture of tachyon and scalar
(phantom) fields without interaction and (v) mixture of tachyon
and scalar (phantom) fields with interaction.\\

{\bf{Case I: GSL in presence of only tachyonic field}}
\\

\begin{figure}
\includegraphics[height=2.5in]{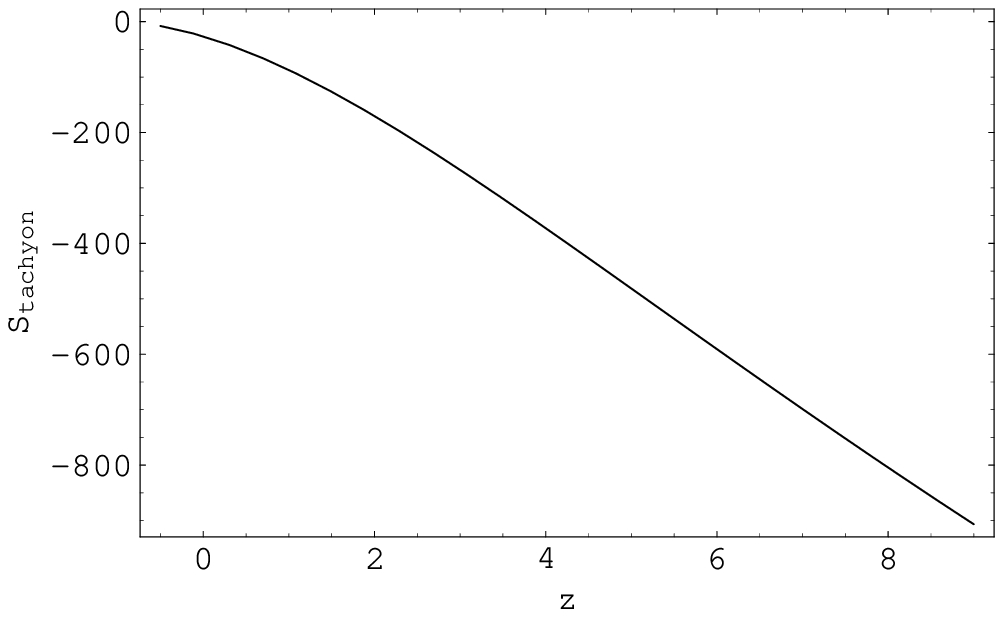}~\\
\vspace{1mm} ~~~~~~~~~~~~Fig.1 \vspace{6mm} shows the plot of
$\dot{S}_{tachyon}$ against $z$ when only tachyonic field is
considered. The vertical axis represents the time derivative of
entropy.

\vspace{6mm}

\end{figure}

We consider equations (4) and (5) where the potential is obtained
from equation (12) with $\delta=0$. Subsequently
$\dot{S}_{tachyon}$ is plotted against redshift $z$ in figure 1.
It is observed from the said figure that  $\dot{S}_{tachyon}$ has
an increasing nature with decrease in $z$. In this figure is it
also visible that $\dot{S}$ is remaining negative when only
tachyonic field is considered. At the same time we have noted that
$R_{h}$ always remains at positive level.
\\

{\bf{Case II: GSL in presence of only phantom field}}
\\

\begin{figure}
\includegraphics[height=2.5in]{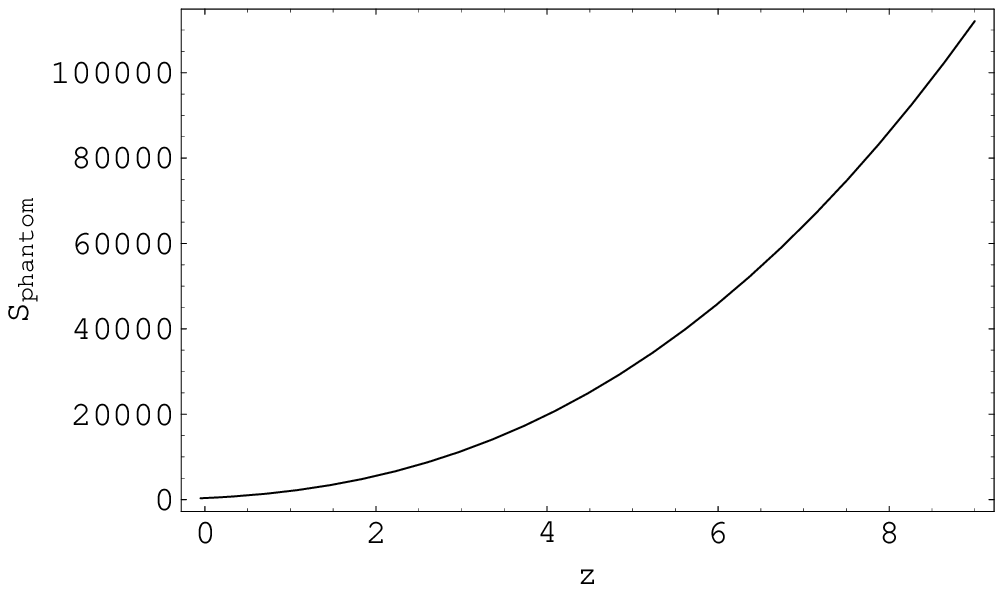}~\\
\vspace{1mm} ~~~~~~~~~~~~Fig.2 \vspace{6mm}  plots
 $\dot{S}_{phantom}$ against $z$ when only phantom field is
considered. The time derivative of entropy is presented along the
vertical axis.

\vspace{6mm} \vspace{6mm}

\end{figure}

We consider equations (6) and (7) with $\epsilon=-1$ and the
potential is obtained from equation (13) with $\delta=0$.
Subsequently $\dot{S}_{phantom}$ is plotted against redshift $z$
in figure 2. It is observed from the said figure that
$\dot{S}_{phantom}$ has a decreasing nature with decrease in $z$.
In this figure is it also apparent that $\dot{S}$ is remaining
positive when only phantom field is considered. At the same time
we have noted that $R_{h}$ always remains at positive level.
\\

{\bf{Case III: GSL in presence of only scalar field}}
\\

\begin{figure}
\includegraphics[height=2.5in]{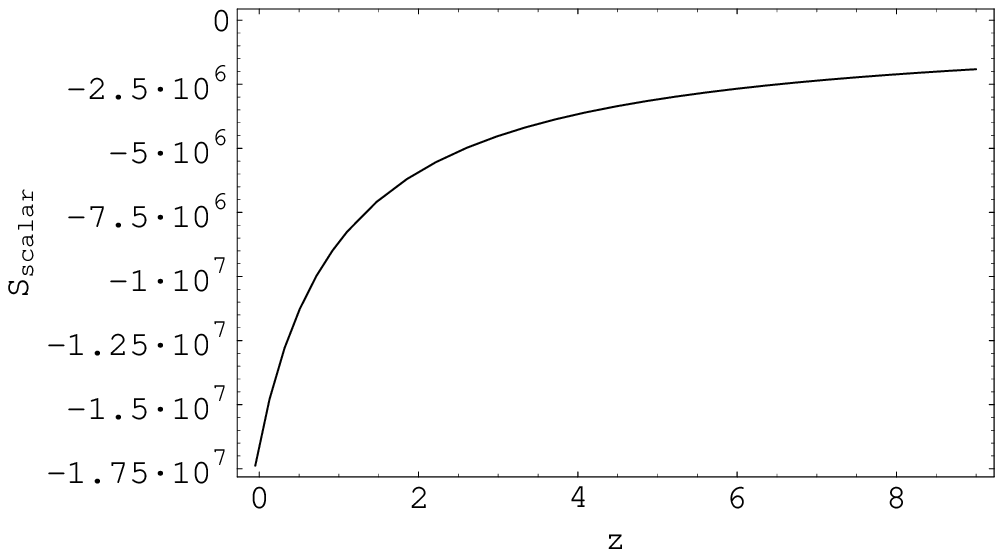}~\\
\vspace{1mm} ~~~~~~~~~~~~Fig.3 \vspace{6mm} plots
$\dot{S}_{scalar}$ against $z$ when only scalar field is
considered. The time derivative of entropy is presented along the
vertical axis.

\vspace{6mm}

\end{figure}

We consider equations (6) and (7) with $\epsilon=1$ and the
potential is obtained from equation (13) with $\delta=0$.
Subsequently $\dot{S}_{scalar}$ is plotted against redshift $z$ in
figure 3. It is observed from the said figure that
$\dot{S}_{scalar}$ has a decreasing nature with decrease in $z$.
In this figure is it also apparent that $\dot{S}$ is remaining
negative when only scalar field is considered. At the same time we
have noted that $R_{h}$ always remains at positive level.
\\

{\bf{Case IV: GSL in presence of a mixture of tachyonic field and
scalar (phantom) field without interaction}}
\\

\begin{figure}
\includegraphics[height=2.0in]{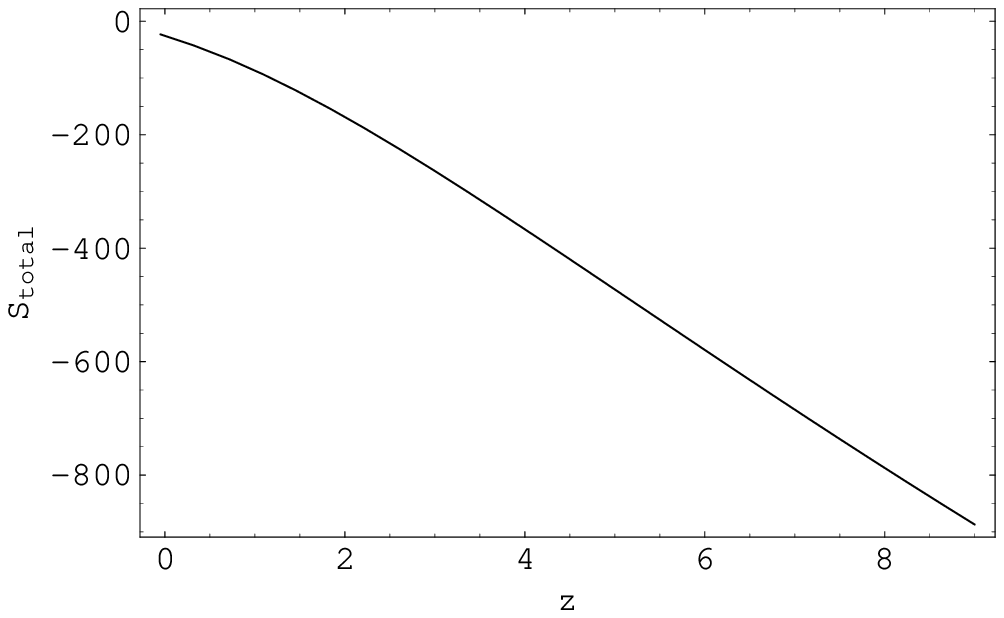}~~~
\includegraphics[height=2.0in]{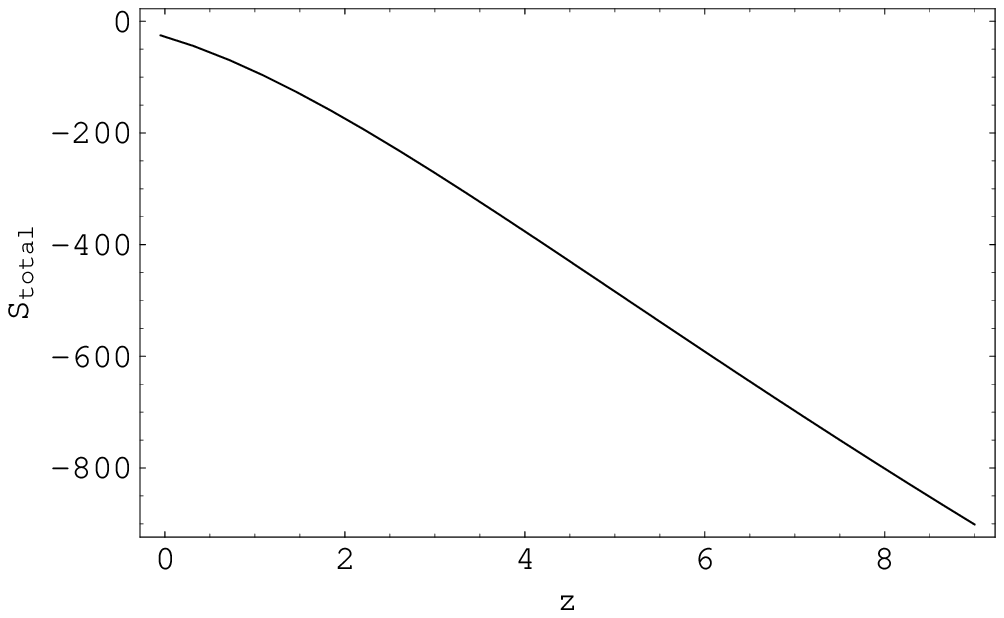}~\\
\vspace{1mm} ~~~~~~~~~~~~Fig.4~~~~~~~~~~~~~~~~~~~~~~~~~~~~~~~~~~~~~~~~~~~~~~~~~~~~~~~~Fig.5\\
\vspace{6mm} Figs. 4 and 5 plot $\dot{S}_{total}$ against $z$ when
a tachyonic field has mixed with phantom field and scalar field
respectively without interaction ($\delta=0$). The time derivative
of entropy is presented along the vertical axes.

\vspace{6mm} \vspace{6mm}

\end{figure}
Equations (2) and (3) are used to calculate the energy density and
pressure of the two-component system of tachyonic field and scalar
(phantom) field. $\rho_{1}$, $\rho_{2}$, $p_{1}$, and $p_{2}$ are
computed using (4)-(7), and (10)-(13). However, the
non-interacting situation being considered, we put $\delta=0$.
Figures 4 and 5 display the evolution of $\dot{S}_{total}$ in the
cases of tachyonic field mixing with phantom and scalar field
respectively. Both of the figures display almost similar pattern
of evolution of $\dot{S}_{total}$ with redshift $z$. In both of
the cases the derivative of the entropy is staying at negative
level and is increasing with decrease in the redshift. In both of
the cases we have seen that the event horizon $R_{h}$ is remaining
positive.
\\

{\bf{Case V: GSL in presence of interaction between tachyonic
field and scalar (phantom) field }}
\\

In the present case, we consider the interacting situation.
Therefore, in the conservation equation (8), the interaction
parameter $\delta$ is kept at non-zero level. Equations (2) and
(3) are used to calculate the energy density and pressure of the
two-component system of tachyonic field and scalar (phantom)
field. $\rho_{1}$, $\rho_{2}$, $p_{1}$, and $p_{2}$ are computed
using (4)-(7), and (10)-(13). Figures 6 and 7 display the
evolution of $\dot{S}_{total}$ in the cases of tachyonic field
interacting with phantom and scalar field respectively. It is
found that the figures 6 and 7 are displaying similar pattern.
That is, in presence of interaction, the variation of
$\dot{S}_{total}$ with decrease in redshift is almost same i.e.
increasing. Moreover, the pattern has similarity with the
non-interacting cases too. Here also, the derivative of the
entropy is remaining at negative level. We have tested that
$R_{h}$ is keeping positive with decrease in the redshift. The
figures display the plotting with $\delta=0.03$. Despite varying
the values of $\delta$ to other positive and negative values the
pattern has remained the same.
\\

\begin{figure}
\includegraphics[height=2.0in]{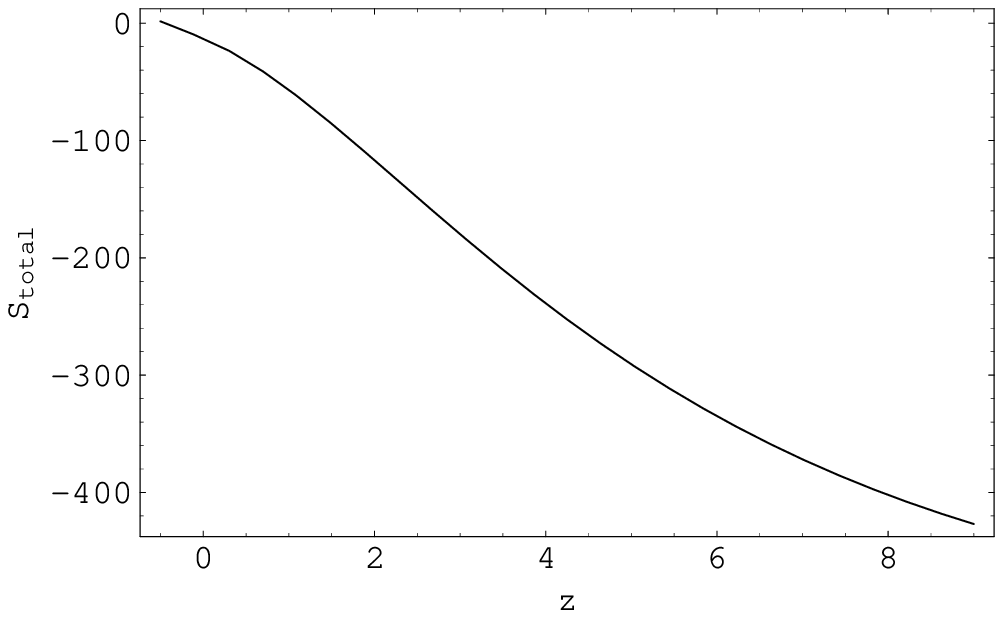}~~~
\includegraphics[height=2.0in]{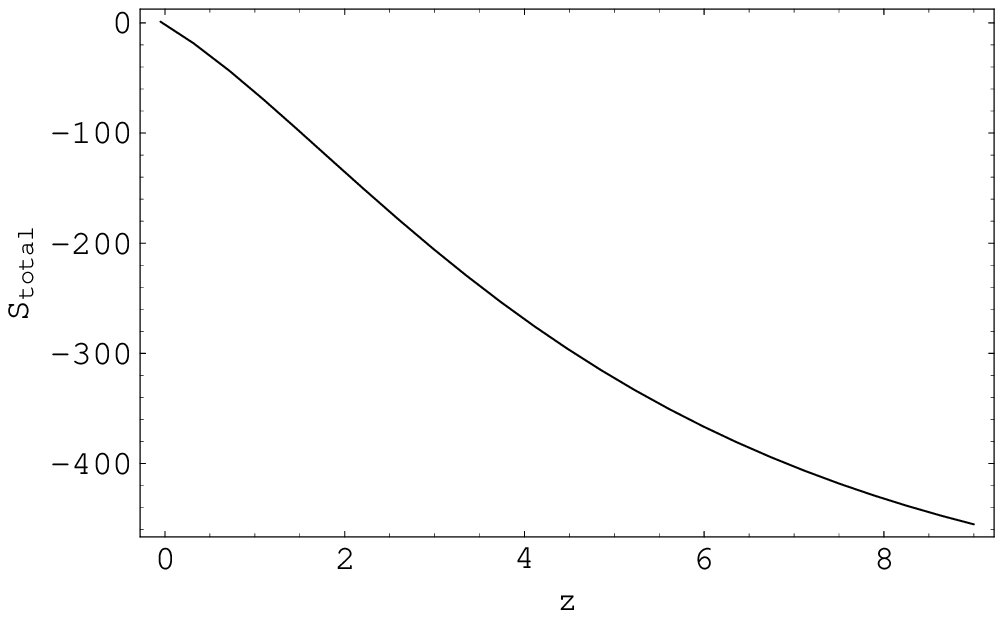}~\\
\vspace{1mm} ~~~~~~~~~~~~Fig.6~~~~~~~~~~~~~~~~~~~~~~~~~~~~~~~~~~~~~~~~~~~~~~~~~~~~~~~~Fig.7\\
\vspace{6mm} Figs. 6 and 7 show the plot of $\dot{S}_{total}$
against $z$ when tachyonic field is interacting with phantom and
scalar field respectively ($\delta=0.03$). The time derivative of
entropy is presented along the vertical axis.

\vspace{6mm}

\end{figure}

In this work, we fixed our target at answering the question:
\emph{Is $\dot{S}>0$ under the interaction between tachyonic field
and scalar (phantom) field}? In the previous paragraphs, we have
searched for the answer to this question. While doing so we have
investigated several competitive cases. The candidates of dark
energy, namely, tachyonic field, phantom field, and scalar field
have been considered in this study. Investigating the second law
of thermodynamics, the globally accepted principle in the
universe, both single component and two-component models
(interacting and non-interacting) have been considered. In the
flat FRW universe, we have seen that in all the above cases the
event horizon $R_{h}$ is staying at positive level. However, the
derivative of the entropy $\dot{S_{X}}=\dot{S}+\dot{S_{h}}$ has
displayed varied natures in the cases under consideration. When we
considered a flat FRW universe dominated by tachyonic field only,
we found that the derivative of the entropy is increasing with
decrease in the redshift. However, in phantom dominated and scalar
field dominated cases, it is decreasing with decrease in the
redshift. Another notable thing is that, excepting the phantom
dominated universe, the derivative of the entropy is remaining at
the negative level. We further noted that, the increase in the
derivative of the entropy in the case of tachyonic field dominated
universe is more sharp than the decrease in the derivative of the
entropy with decrease in the redshift in the other two cases. We
have further seen that if we consider the two-component model,
then also the evolution of the derivative of the entropy with
decrease in the redshift is almost similar to the tachyonic field
dominated universe. Even the introduction of interaction parameter
$\delta$ at positive or negative level does not affect this
pattern significantly. In these cases too, the derivative of the
entropy remains negative.
\\\\
{\bf Acknowledgement:}\\
The authors wish to acknowledge the warm hospitality provided by
the Inter University Centre for Astronomy and Astrophysics
(IUCAA), Pune, India, where the work was carried out during a
scientific visit in January, 2010. The authors are thankful to the
anonymous reviewers for giving constructive comments to enhance
the quality of the work.
\\\\

{\bf References:}\\
\\
$[1]$ E. J. Copeland, M. Sami and S. Tsujikwa, {\it IJMPD} {\bf
15}, 1753 (2006).\\
$[2]$A. G. Riess et al., {\it Astrophys. J.} {\bf
116}, 1009 (1998).\\
$[3]$ A. G. Riess et al., {\it Astron. J.} {\bf 117},
707 (1999).\\
$[4]$ Y.-F. Cai, E. N. Saridakis, M. R. Setare and J.-Q. Xia, {\it Phys. Rept.} {\bf 493}, 1 (2010).\\
$[5]$ S. J. Perlmutter et al, {\it Astrophys. J.} {\bf 517}, 565
(1999).\\
$[6]$ R. R. Caldwell, {\it Phys. Lett. B} {\bf 545}, 23 (2002).\\
$[7]$ A. Sen, {\it JHEP} {\bf 0204} 048 (2002); {\it JHEP} {\bf 0207}, 065 (2002).\\
$[8]$ G. W. Gibbons, {\it Phys. Lett. B} {\bf 537}, 1 (2002).\\
$[9]$ M. Sami, {\it Mod. Phys. Lett. A} {\bf 18}, 691 (2003).\\
$[10]$ M. S. Berger and H. Shojaei, {\it Phys. Rev. D} {\bf 74},
043530 (2006).\\
$[11]$ R. Herrera, D. Pavon and W. Zimdahl, {\it Gen. Rel. Grav.}
{\bf 36}, 2161 (2004).\\
$[12]$ R. -G. Cai and A. Wang, {\it JCAP} {\bf 0503}, 002 (2005).\\
$[13]$ Z. -K. Guo, R. -G. Cai and Y. -Z. Zhang, {\it JCAP} {\bf 0505}, 002 (2005).\\
$[14]$ T. Gonzalez and I. Quiros, {\it Class. Quantum Grav.} {\bf 25}, 175019 (2008).\\
$[15]$ G. Izquierdo and D. Pavon, {\it Phys. Lett. B} {\bf 633}, 420 (2006).\\
$[16]$ M. R. Setare, {\it Phys. Lett. B} {\bf 641}, 130 (2006).\\
$[17]$ M. R. Setare and S. Shafei, {\it JCAP} {\bf 09}, 011 (2006)\\
$[18]$ M. R. Setare, {\it JCAP} {\bf 01}, 023 (2007).\\
$[19]$ M. Sami, P. Chingangbam and T. Qureshi, {\it Pramana} {\bf
62}, 765 (2004).\\
$[20]$ S. Chattopadhyay, U. Debnath and G. Chattopadhyay, {\it
Astrophys. Space Sci.} {\bf 314}, 41 (2008).\\
$[21]$ P. Singh, M. Sami and N. Dadhich, {\it Phys. Rev. D} {\bf
68}, 023522 (2003).\\
$[22]$ A. de la Macorra, {\it JCAP} 030 (2008).\\
$[23]$ S. Chattopadhyay and U. Debnath, {\it Brazilian J. Phys.} {\bf 39}, 86 (2009).\\
$[24]$ M. Cataldo, P. Mella,  P. Minning and J. Saavedra, {\it
Phys. Lett. B} {\bf 662}, 314 (2008).\\
$[25]$ M. R. Setare, J. Sadeghi and A. R. Amani, {\it Phys. Lett.
B} {\bf 673}, 241 (2009).\\
$[26]$ B. Wang, Y. Gong and E. Abdalla, {\it Phys. Lett. B} {\bf
624}, 141 (2005).\\
$[27]$ M. R. Setare , {\it Eur. Phys. J. C} {\bf 50}, 991 (2007).\\

\end{document}